# Tailoring π-conjugation and vibrational modes to steer on-surface synthesis of pentalene-bridged ladder polymers

Bruno de la Torre[a,b]♦, Adam Matěj[a,b]♦, Ana Sánchez-Grande[c]♦, Borja Cirera[c], Benjamin Mallada[a,b], Eider Rodríguez-Sánchez[c], José Santos[c,d], Jesús I. Mendieta-Moreno [b], Shayan Edalatmanesh[a,b], Koen Lauwaet[c], Michal Otyepka[a], Miroslav Medveď[a], Álvaro Buendía[e], Rodolfo Miranda[c,e], Nazario Martín*[c,d], Pavel Jelínek*[a,b] and David Écija*[c]

[a]Regional Centre of Advanced Technologies and Materials, Palacký University, Šlechtitelů 27, 78371 Olomouc, Czech Republic.
[b]Institute of Physics, The Czech Academy of Sciences. Cukrovarnická 10, 162 00 Prague 6, Czech Republic.
[c]IMDEA Nanociencia, C/ Faraday 9, Ciudad Universitaria de Cantoblanco, 28049, Madrid, Spain.
[d]Departamento de Química Orgánica, Facultad de Ciencias Químicas, Universidad Complutense, 28040 Madrid, Spain.
[e]Departamento de Física de la Materia Condensada, Universidad Autónoma de Madrid, Cantoblanco, Madrid, Spain.

Corresponding authors:
nazmar@ucm.es, jelinekp@fzu.cz, david.ecija@imdea.org

♦ Equally contributing authors

ABSTRACT

The development of synthetic strategies to engineer π-conjugated polymers is of paramount importance in modern chemistry and materials science. Here we introduce a theoretical and experimental synthetic paradigm based on the search for specific vibrational modes through an appropriate tailoring of the π-conjugation of the precursors, in order to increase the attempt frequency of a chemical reaction. First, we on-surface design a 1D π-conjugated polymer with specific π-topology, which is based on bisanthene monomers linked by cumulene bridges that tune specific vibrational modes. In a second step, upon further annealing, such vibrational modes steer the two-fold cyclization reaction between adjacent bisanthene moieties, which gives rise to a long and free-defect pentalene-bridged conjugated ladder polymer featuring a low band gap. In addition, high resolution atomic force microscopy allows us to identify by atomistic insights the resonant form of the polymer, thus confirming the validity of the Glidewell and Lloyd´s rules for aromaticity. This on-surface synthetic strategy may stimulate exploiting previously precluded reactions towards novel π-conjugated polymers with specific structures and properties.

INTRODUCTION

The design and synthesis of π-conjugated polymers is a very active area of research with great potential application in organic field-effect transistors (OFETs), photovoltaics (OPVs) and light emitting diodes (OLEDs)[1,2,3].

However, the advance in the synthesis of new π-conjugated polymers is hampered by concomitant limitations of solubility during the chemical synthesis[2,3]. This drawback is even enhanced in the design of π-conjugated ladder polymers, a singular type of polymers in which all the backbone is π-conjugated and fused[4]. These conjugated ladder polymers are of great appeal for materials science and optoelectronics due to their exceptional stability and optimum electron delocalization thanks to the restriction of free torsional motion in between monomers[4].

In the field of π-conjugated polymers, the topology of the π-electron network is crucial since it determines the ground state electronic structure of such materials. Polymers incorporating non-benzenoid polycyclic hydrocarbons are of increasing interest due their specific electronic properties[5]. In non-benzenoid systems, molecular orbital levels and π-electron density distribution are uneven compared to benzenoid systems, thereby polarizing the ground state and leading to unique behavior in excited states[5]. π-Extended non-benzenoid polycyclic hydrocarbons, such as pentalene, indacene or indenofluorene, have recently propelled rich insights into the electronic properties of antiaromaticity and/or open-shell character[6,7,8,9,10,11,12]. On the one hand, antiaromatic compounds show better conducting properties than their aromatic counterparts[13]. On the other hand, non-benzenoid species are prone to express open-shell character in solution[6,7] and on surfaces[9,10,11,12].

Taking into account the aforementioned unique properties, it is of timely relevance to engineer conjugated ladder polymers incorporating non-benzenoid components, targeting to design chemically robust and low bandgap polymers. Such synthesis in wet chemistry must contend with structural defects and low solubility that prevent complete control over the synthesis and structural characterization at the atomic scale. To overcome these synthetic challenges, on-surface synthesis has emerged as a disruptive paradigm to develop chemical reactions on surfaces, while simultaneously monitoring and elucidating the precursors, intermediates and reaction products by means of scanning probe microscopy[14,15,16,17,18]. Despite the recent progress in on-surface synthesis, there are still very limited strategies to synthesize complex π-conjugated polymers with specific properties[11,15,19,20,21].

The electronic and chemical properties of π-conjugated polymers are driven by the inherent topology of their π-electrons. Here, we present a chemical strategy to allow unconventional synthetic pathways by exploiting the relation between π-conjugation and specific vibrational modes, being able to steer a desired chemical transformation. Our strategy is based on two sequential steps, i.e. a polymerization expressing specifically tailored π-topology, and a subsequent complex chemical ladderization involving a two-fold cyclization by taking advantage of the unique vibrational capabilities of the inherent π-topology. In the first step, the on-surface formation of a one-dimensional polymeric precursor featuring cumulene bridges takes place, allowing bending vibrational modes of the polymer with a desired directionality. In the second and final steps, subsequent thermal annealing drives intrapolymeric cyclization reactions, triggered by such specific vibrational modes, giving rise to ladderization of the polymer and the expression within the polymeric main-chain of non-benzenoid fused pentalene bridges.

## RESULTS AND DISCUSSION

Following this chemical strategy, we report the synthesis and detailed atomistic characterization by a comprehensive scanning tunneling microscopy (STM) and non-contact atomic force microscopy (nc-AFM) supported by density functional theory (DFT) study of the on-surface reactions of two distinct molecular precursors, **4BrBiA** [10,10'-bis(dibromomethylene)-10*H*,10'*H*-9,9'-bianthracenylidene] and **4BrAn** [9,10-bis(dibromomethylene)-9,10-dihydroanthracene]. The precursors are endowed with =CBr$_2$ functionalities to allow homocoupling[19,20,21]. The deposition of **4BrAn** and **4BrBiA** on Au(111) and subsequent annealing at 500 K results in the formation of anthracene (**1**, Fig.1a) and bisanthene (**2**, Fig. 1b) polymers, respectively[19,20,21]. Both polymers exhibit repeating moieties linked by linear bridges, but with distinct π-conjugation. While the anthracene polymer adopts aromatic ethynylene π-conjugation character[19], the bisanthene polymer results in the promotion of the quinoid cumulene-like character[20,21].

Further annealing at 650 K leads to distinct behavior of the formed polymers despite their chemical similarity. Our theoretical calculations (see later) reveal that, as a result of this different π-topology, the cumulene-like bridge (with a weakened triple bond) in the bisanthene polymer **(2)** allows the amplification of two bending vibrational modes of the bridging unit (see Fig. 2a), which promotes a distinct reaction mechanism with respect to the anthracene polymer **(1)** (see Fig. 1a and Supplementary Fig. 1). In the case of the cumulene-bridged bisanthene polymer **(2)** (see Fig. 1c) these vibrational modes substantially overlap with the intrapolymeric reaction coordinate. Their thermal excitation upon annealing at 650 K promotes the formation of fused pentalene bridges, which gives rise to a long and defect-free π-conjugated ladder polymer **(3)**, as shown in Figure 1d. High-resolution nc-AFM images acquired with CO-tip reveal that the fused bridging units are formed by pentalene moieties. Notably, this intrapolymeric ladderization reaction is 100% selective and no side-reactions are detected for submonolayer coverage. On one hand, for the case of the anthracene polymer (**1**), annealing above 700 K results in distinct side reactions giving rise to irregular polymers with variety of defects (see Supplementary Fig. 1). Thus, the annealing of cumulene-linked bisanthene polymers allows new chemical reactions that are precluded for the ethynylene-bridged anthracene wires.

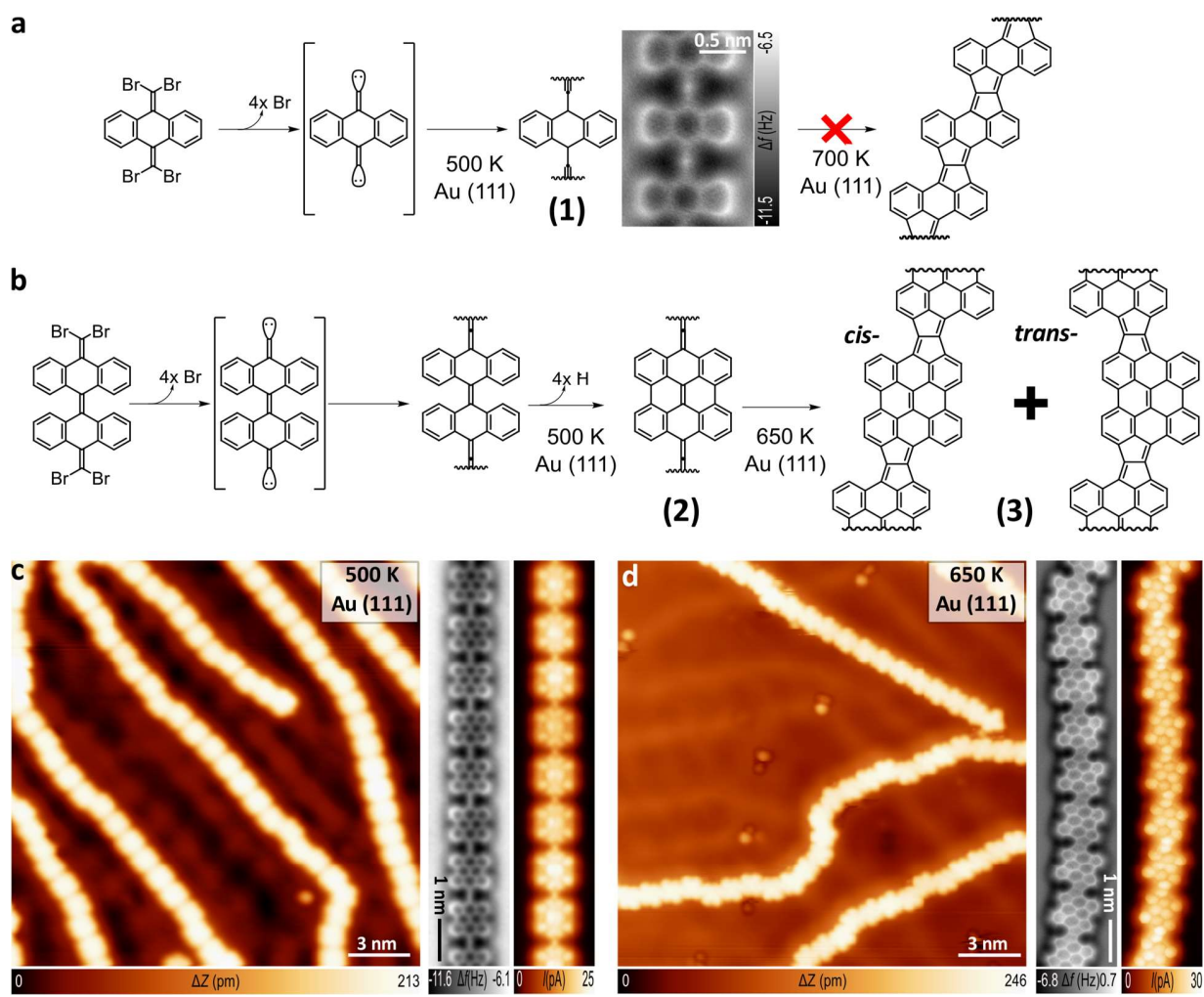

**Fig. 1 | On-surface synthesis of pentalene-bridged bisanthene polymers. a,** Scheme of the reaction sequence of **4BrAn** precursor after being deposited on Au(111), annealed to 500 K to obtain polymer **1**, and annealed up to 700 K to only produce irregular defective polymers (see Supplemental Fig. 1). **b,** Scheme of the reaction sequence of **4BrBiA** precursor after being deposited on Au(111), annealed to 500 K to obtain polymer **2**, and annealed to 650 K to result in polymer **3**. **c,** STM topographic overview (left panel; 0.03 V, 0.1 nA) and detailed nc-AFM/STM constant height images (right panels) of cumulene-linked bisanthene polymers (**2**). **d,** STM topographic overview (left panel; 0.05 V, 0.1 nA) and detailed nc-AFM/STM constant height images (right panels) of pentalene-bridged bisanthene polymers (**3**).

To get a deeper insight into these differences and the underlying reaction mechanism, we carried out QM/MM simulations[22,23] with the aim of analysing the optimal reaction pathways of the bridging units of both anthracene and bisanthene polymers towards the formation of pentalene links. In general, the feasibility of a chemical reaction is determined by two factors: i) the Boltzmann exponential factor given by the activation energy barrier, and ii) the attempt frequency $\nu$ expressing how often the system evolves towards the saddle point of the reaction pathway[24,25,26,27]. In the following, we will show that the π-conjugation form adopted by the polymer may also strongly influence the optimal reaction pathway by tuning both the activation energy and steering vibrational modes.

To find the most feasible reaction pathway towards the formation of the pentalene bridged polymer (**3**) (see Supplementary Fig. S3), we attempted different reaction coordinates mimicking the first step of the reaction. We have considered either a dissociation of the C-H bond at the zigzag edge of the bisanthene monomer mediated by an Au adatom (see Fig. 2) or a direct cyclization between the bridges and the rims of the monomers (see Supplementary Fig. S3). The QM/MM calculations at 650 K show that the activation free energy for hydrogen abstraction (34 and 37 kcal/mol) is much lower than that of the direct cyclization (82 and 110 kcal/mol) for both bisanthene (**2**) and anthracene (**1**) polymers, respectively (see Supplemental Figs. S3 and S4). Thus, we assume that the first step of the reaction pathway is a hydrogen abstraction in both cases.

In the second step of the reaction, the cyclization process for both polymers (**1** and **2**) have slightly lower activation barrier than a second dehydrogenation by a few kcal/mol (see Supplementary Figs. S3 and S4). The calculated activation free energy of cyclization for the bisanthene polymer (**2**) is significantly smaller (23 kcal/mol) than that for the anthracene polymer (**1**, 32 kcal/mol). As the activation barriers of the cyclization process in the second reaction step are lower than those corresponding to the second hydrogenation abstraction, the dehydrogenation-cyclization-dehydrogenation scenario (shown in Fig. 2 and also denoted by yellow lines in Supplementary Figs. 3 and 4) seems to be the optimal reaction pathway among the explored reaction pathways for both polymers. Nevertheless, experimentally we only observe the formation of the long and defect-free pentalene bridged polymers from the bisanthene (**2**), but not from the anthracene (**1**) polymers. Thus, the activation energies cannot explain by themselves the experimental evidence.

However, the reaction rate also depends on the $\nu$ prefactor, which is determined by available steering vibrational eigenmodes strongly overlapping with the reaction coordinates. Figure 2a,b shows the atomic arrangement in the transition state (shown in transparent mode) of the cyclization reaction, which involves a bending distortion of the cumulene bridge. According to our analysis of the vibrational modes of bisanthene polymer (**2**), we can identify two bending vibrational modes that are coupled with the in-phase displacement of the unsaturated carbon atom of the bisanthene moiety (see red arrows in Fig. 2a). Such bending vibrational modes match very well with the cyclization reaction coordinate depicted by blue arrows in Fig. 2a (also see vibrational modes of bisanthene labeled 182 and 185 shown in Supplementary Fig. S5).

On the contrary, in the case of the anthracene polymer (**1**), all the vibrational modes consisting of the bending mode lack coupling with the in-phase movement of the unsaturated carbon atom (see Fig. 2b). The anthracene polymer (**1**) contains the ethynylene bridges, thus featuring a triple bond, that has distinct bonding character than the cumulene-like linkers of bisanthene polymer (**2**)[20,21]. Consequently, the frequencies of the steering bending modes of the ethynylene linkers in anthracene polymers (**1**) have lower frequencies by 50-100 cm$^{-1}$ than those of cumulene-like bridges in bisanthene polymers (**2**), as shown in Supplementary Fig. S5. Therefore, in the case of the cumulene-like bridged bisanthene polymer (**2**) the presence of the bending vibrational modes that strongly overlap with the cyclization reaction coordinates together with slightly higher vibrational frequencies provide larger attempt frequency $\nu$ than in the case of ethynylene-linked anthracene polymer (**1**).

In summary, the distinct π-conjugation of the bisanthene polymer (**2**) exhibiting the cumulene-like bridge is prone to the bending vibrational modes that steer the cyclization reaction. These bending vibrational modes coupled with the appropriated in-phase motion of bisanthene moiety enhance not only the attempt frequency $\nu$, but they also impose a lower energy cost of the bond distortion contributing to the activation energy than for the anthracene polymer (**1**). Altogether, these two factors facilitate the formation of the defect-free pentalene-bridged bisanthene polymers (**3**) as observed experimentally.

At this point, it is worth to point out that the hydrogenation abstraction is driven by the C-H stretching mode (~3100 cm$^{-1}$), which is approximately one order of magnitude larger than the frequencies of the C-C bending modes (~600 cm$^{-1}$), see Supplemental Fig. S5. Thus, the $\nu$ prefactor and, consequently, the total rate of the dehydrogenation processes are enhanced. In the case of anthracene polymer (**1**), the energy barrier (36 kcal/mol) for the second dissociation is only slightly higher than that of the cyclization (31 kcal/mol) (see Supplemental Fig. S4). Consequently, considering that the steering C-H stretching mode is approximately one order of magnitude higher than the C-C bending modes, the reaction rates for the second dissociation or the cyclization at 700 K become similar. This opens a second reaction channel that includes a subsequent double dehydrogenation and cyclization reaction trajectory (denoted as a blue line in Supplemental Fig. S4). Notably, for this reaction trajectory of the anthracene polymer (**1**), the formation of the pentalene bridge has very large activation energy (>100 kcal/mol). Thus, a formation of other defective structures is preferred, which is experimentally confirmed as illustrated in

Supplementary Fig. S1. Such findings mean that other reaction pathways, including the double H-dissociation and cyclization, with lower activation energies and/or suitable active vibrational modes matching corresponding reaction coordinates, take place avoiding the formation of the anthracene pentalene-bridged polymers.

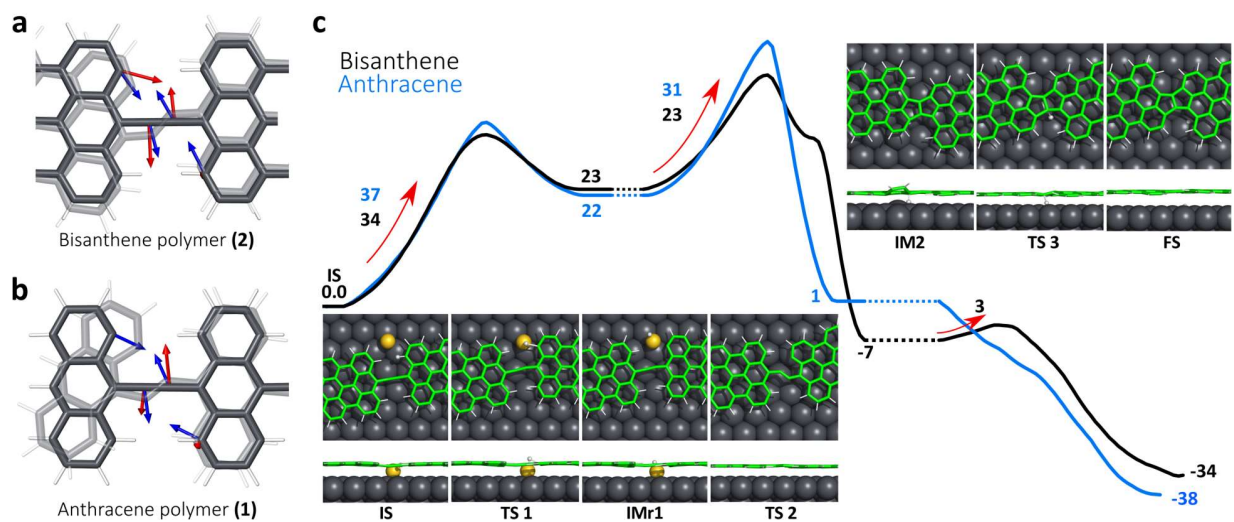

**Fig. 2 | Calculated vibrational modes and reaction pathways of polymers (1) and (2) to give rise to pentalene bridges on Au(111). a, b,** Comparison of calculated vibrational mode (red arrows) with reaction coordinates (blue arrows) for **a,** cumulene-like bridged bisanthene polymer (**2**) and for **b,** ethynylene-linked anthracene polymer (**1**). The atomic arrangement of the transition state (**TS2**) is showed in transparent mode. Reaction coordinates are defined as the displacement from the initial (**IMr1**) to the transition state (**TS2**) for the cyclization reaction pathway from **IMr1** to **IM2**. **c,** Free energy barriers (in Kcal/mol)for the optimal reaction pathway calculated for anthracene (**1**, blue) and bisanthene (**2**, black) polymers at 650 K. Inset figures show top and side views of the initial state (**IS**), transition states (**TS**), intermediates (**IM**) and final states (**FS**) of bisanthene polymer (**2**) optimal reaction pathway towards pentalene-bridged bisanthene polymer (**3**).

Next, we focus on the structural and the electronic properties of new pentalene-bridged bisanthene polymers (**3**), including the dominant resonant form. In general, the bisanthene-pentalene-bisanthene connections can feature *trans* or *cis* (Fig. 1b) configurations, in which the bisanthene moieties are on or off the vertical axis of symmetry with respect to each other, respectively. A statistical analysis of the abundance of the covalent motifs reveals that both configurations are equally favourable. However, segments having repeating *trans* or *cis* motifs were found with a maximum length of twelve units.

Concerning the aromaticity, from a chemical point of view, since there are non-benzenoid units, the aromatic rule to be used here is the one proposed by Glidewell and Lloyd[28,29]. This rule states that the total population of π-electrons in conjugated polycyclic systems tends to form the smallest 4n+2 groups and to avoid the formation of the smallest 4n groups. The pentalene bridge can show two types of resonant forms, i.e. the inner bond can be single or double. From this assumption, at least two distinct resonant forms can be plotted as illustrated in Figure 3a. By applying the Glidewell and Lloyd´s rule, it is evident that the conjugated structure associated with an inner single bond (black structure in Figure 3a) should be the most stable, since the other option (grey structure in Figure 3a) would imply the formation of 4 groups with 4π-electrons in the bisanthene moiety, which must be avoided according to the rule. Interestingly, by following the observed π-conjugation, the system stabilizes four Clar´s sextets in the bisanthene, the maximum number.

To assess the expression of the Glidewell and Lloyd´s rule on surfaces, we take advantage of the capabilities of nc-AFM with functionalized CO-tip[30,31] to resolve the C-C bond order[32]. High resolution nc-AFM images of the pentalene-bridged polymers (see Figure 3b) clearly show distinct bond lengths within the inner part of the polymer, whose statistically average value is plotted in the central panel of Figure 3b (see Supplementary Fig. S2 for more details). The variation of bond distances is qualitatively confirmed by DFT calculations (right panel of Figure 3b). The results reveal that the variation in the bond length of the polymer bonds matches the π-resonant form predicted by the Glidewell and Lloyd´s rule, confirming its validity.

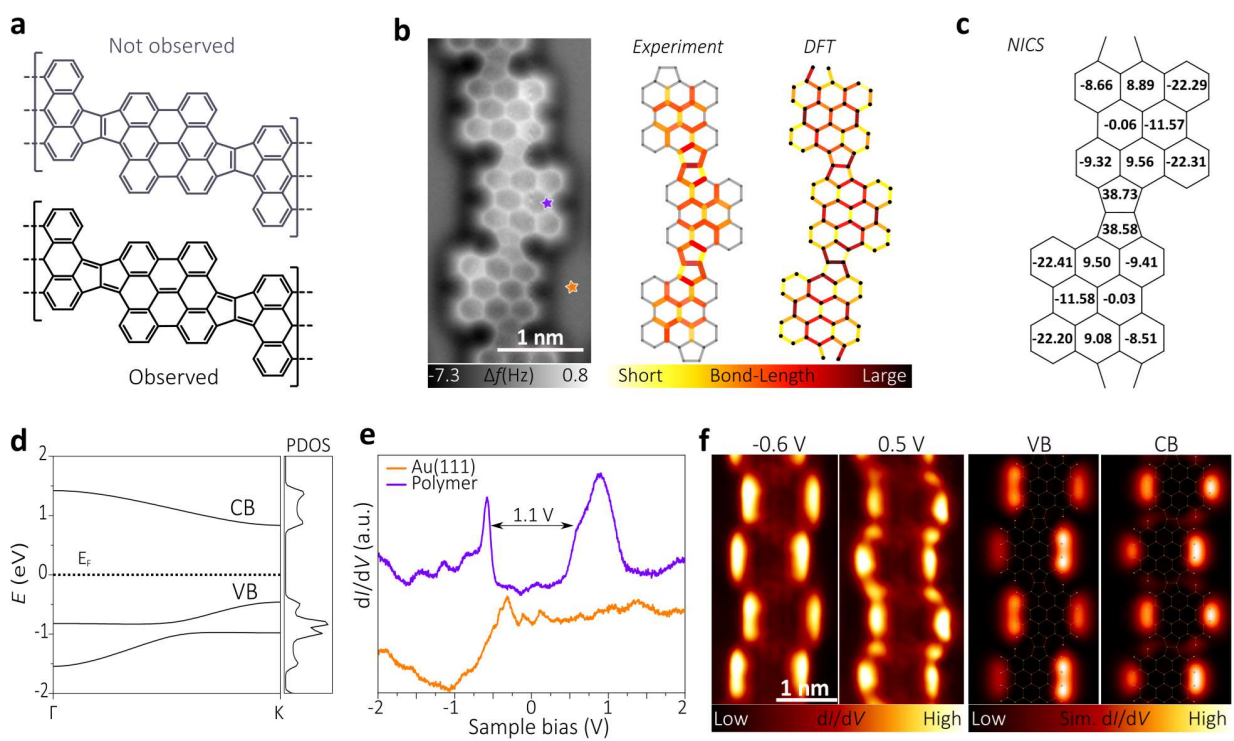

**Fig. 3 | Resonant and electronic structure of the on-surface synthesized pentalene-bridged bisanthene polymer (3). a**, Not observed and observed resonant forms of polymer **3**, featuring *trans* configuration. **b**, High resolution nc-AFM image of a section of the pentalene-bridged bisanthene polymer (left panel) and sketch illustrating the average bond lengths statistically measured from experimental nc-AFM images (central panel) and DFT calculation (right panel) of the bond lengths of a *trans*- section of the polymer. **c**, NICS calculations of a pentalene-linked bisanthene dimer at distance of 1 Å above the center of each ring. **d,** Calculated electronic structure of a free-standing cumulene-like bridged bisanthene polymer (**3**). **e,** dI/dV scanning tunneling spectra acquired at selected site on polymer **3** and on Au(111) as indicated by stars in **b**. **f,** Experimental and simulated dI/dV maps of the valence (VB) and conduction band (CB) of polymer **3.**

Next, the degree of aromaticity and antiaromaticity of the bisanthene polymer is characterized by performing nucleus-independent chemical shift calculations (NICS). Positive values of NICS indicate antiaromaticity, close-to-zero values reveals non-aromaticity, and negative figures correspond to aromaticity. Figure 3c shows that the five-membered rings are highly antiaromatic, whereas the outer six-membered rings of the bisanthene moieties are aromatic, thus corroborating the resonant form discussed above.

In order to elucidate the electronic structure of the π-conjugated polymers, we first calculated the electronic structure of freestanding polymer using B3LYP-DFT, which revealed the presence of dispersive valence (VB) and conduction (CB) bands separated by a band gap of ~ 1 eV (Figure 3d). Spatially resolved d$I$/d$V$ scanning tunneling spectra were recorded over the molecular wires and the clean Au(111) surface. As illustrated in Figure 3e, one frontier resonance is distinguished at -0.6 eV, assigned to the valence band (VB) edge and a broad peak with an elbow at 0.5 eV, interpreted as the conduction band (CB) edge, displaying a maximum at 0.9 eV. This results in a low bandgap of ~1.1 eV. We should note that the bandgap value obtained from STS measurements is typically reduced by an additional electron screening imposed by the proximity of a metallic surface with respect to the intrinsic band gap of the gas phase polymer[33,34]. Within the bandgap the downshifted surface state convoluted with tip states is observed. The d$I$/d$V$ map at the VB edge shows maxima over lateral termini of the bisanthene moiety, without charge density over the pentalene bridge (see Figure 3f). The d$I$/d$V$ map of CB edge exhibits states over the armchair region of the bisanthene units and on the void spaces adjacent to the pentalene bridges (see Figure 3f). Despite the fact that DFT simulations cannot predict correctly the magnitude of the intrinsic band gap of the polymer[35], they describe the character of frontier orbitals of the VB and CB edges of the bisanthene polymer very well. Indeed, the agreement between experimental and simulated d$I$/d$V$ maps is excellent, validating the character of the frontier orbitals predicted by DFT.

Thus, the fused pentalene-bridged bisanthene polymers are low bandgap polymers, featuring antiaromatic bridging units, heralding potential for near-infrared activity, high conductivity, and ambipolar charge transport[36,37,38].

CONCLUSIONS

Our results introduce the importance of tailoring π-conjugation and vibrational modes on surfaces to promote otherwise precluded chemical reaction pathways. Following this strategy, we reveal that a cumulene-like bridged bisanthene polymer is prone to undergo ladderization into a fused pentalene-linked bisanthene polymer through a two-fold cyclization process, thanks to specific steering vibrational modes. The absence of such modes in an ethynylene-linked anthracene polymer featuring similar coordination environment, but very distinct π-topology of the linker, blocks the ladderization reaction and reveals the hidden and relevant influence of the resonant form of the bridge in the chemical reaction.

We envision our studies will open avenues to engineer highly demanded conjugated nanomaterials, while showing strategies to incorporate non-benzenoid moieties in polymeric science in order to steer relevant electronic phenomena of interest for molecular optoelectronics and organic solar cells. Notably, our results highlight the versatility of on-surface synthesis for designing non-benzenoid based nanomaterials featuring low bandgaps, which are of utmost interest because of their near-infrared activity and ambipolar charge transport character[36,37,38].

EXPERIMENTAL AND THEORETICAL METHODS

Experiments were performed in two independent custom designed ultra-high vacuum systems that host a low-temperature Omicron and a Createc scanning tunneling microscope, respectively, where the base pressure was below $5\times10^{-10}$ mbar. STM images were acquired with electrochemically etched tungsten tips or cut and sharpened by focus ion beam (FIB) Pt/Ir tips, applying a bias ($V_b$) to the sample at a temperature of ~4 K. Precursor molecules were synthesized in our group following a procedure described in Refs. 19, 20 and 21. The Au(111) substrate was prepared by standard cycles of $Ar^+$ sputtering (800 eV) and subsequent annealing at 723 K for 10 minutes. Molecular precursor were deposited by organic molecular-beam epitaxy (OMBE) from a quartz crucible maintained at 373 K (**4BrAn**) or 443 K (**4BrBiAn**) onto a clean Au(111) held at room temperature. Whenever necessary samples were annealed to the desired temperature and subsequently transferred to the STM stage, which was maintained at 4.2 K. For the spectroscopic measurements, specific site dI/dV and dI/dV maps were acquired with conventional lock-in technique with a modulation of 5 mV and 10 mV respectively. In nc-AFM imaging, a Pt/Ir tip mounted on a qplus sensor (resonant frequency ≈ 30 kHz; stiffness ≈ 1800 N/m) was oscillated with a constant amplitude of 50 pm. The tip apex was functionalized with a CO-molecule, and all images were captured in constant height mode and a bias sample of 1 mV. AFM images are raw data whereas STM images were subject to standard processes. The images were processed using WSxM[39] software.

Density Functional Theory calculations for all freestanding finite and infinite systems (anthracene, pentacene and bisanthene) are done using FHI-AIMS, Fireball packages[40]. All geometry optimizations and electronic structure analyses have been performed using B3LYP[41] exchange-correlation functionals. Systems were allowed to relax until the remaining atomic forces reached below $10^{-2}$ eV/Å. For all infinite systems with periodic boundary condition (PBC), a Monkhorst-Pack grid of 18x1x1 was used to sample the Brillouin zone. Theoretical dI/dV maps were calculated by FHI-AIMS with Probe Particle SPM code[42] a s-like orbital tip[43]. We used a QM/MM method, Fireball/Amber[23] based on the combination of classical force field techniques with Amber[44] and local orbital DFT with Fireball[22]. In the MM part we used the interface force-field[45] and in the DFT calculations we used the BLYP exchange correlation functional[46] with D3 corrections[47] and norm-conserving pseudopotentials. We employed a basis set of optimized numerical atomic-like orbitals (NAOs)[48] with a 1s orbital for H and sp$_3$ orbitals for C atoms. Before the calculation of the free energy profile we have perform a QM/MM geometry relaxation followed by a thermalization of the system from 100K to 500K in order to stabilize the system. The free energy profile was performed with the WHAM method[49]. In each window we run a QM/MM MD of 2000 steps at 500 K with a time step of 0.5 fs. NICS(1)$_{zz}$ values were evaluated using GIAO method[50] at spin-unrestricted ωB97X-D[51]/cc-pVDZ[52] level of theory in Gaussian09[53]. Tetrameric molecules were relaxed at the same level prior to NMR calculations.

ASSOCIATED CONTENT

**Supporting Information**.

AUTHOR INFORMATION

**Corresponding Author**

nazmar@ucm.es, jelinekp@fzu.cz, david.ecija@imdea.org

**Author Contributions**

N.M., P.J. and D.E. conceived and designed the experiments. P.J., N.M, and D.E. supervised the project and led the collaboration efforts. B.T., A.S-G., B.C, B.M. and D.E. carried out the experiments and obtained the data. J.S., E.R.-S. and N. M. synthesized the precursors. The experimental data were analyzed by B.T., A. M., A.S.-G., B.C., K.L., A. B., R.M., P. J. and D.E., and discussed by all the authors. A. M., J.M., S. E., M. O., M. M. and P.J. performed the calculations. The manuscript was written by B.T., N. M., P. J., and D. E., with contribution from all the authors.

**Funding Sources**

Work supported by the Comunidad de Madrid [project QUIMTRONIC-CM (Y2018/NMT-4783)] and the ERC Consolidator Grant ELECNANO (nº 766555). MINECO of Spain (projects CTQ2017-83531-R and RED2018-102815-T) is also acknowledged. IMDEA Nanociencia thanks support from the "Severo Ochoa" Programme for Centers of Excellence in R&D (MINECO, Grant SEV-2016-0686). We also acknowledge support from Praemium Academie of the Academy of Science of the Czech Republic, GACR 18-09914S, 20-13692X and Operational Programme Research, Development and Education financed by European Structural and Investment Funds and the Czech Ministry of Education, Youth and Sports (Project No. CZ.02.1.01/0.0/0.0/16_019/0000754).

# Supplementary Materials for

# Tailoring π-conjugation and vibrational modes to steer on-surface synthesis of pentalene-bridged ladder polymers

**Bruno de la Torre[a,b]♦, Adam Matěj[a,b] ♦, Ana Sánchez-Grande[c]♦, Borja Cirera[c], Benjamin Mallada[a,b], Eider Rodríguez-Sánchez[c], José Santos[c,d], Jesús I. Mendieta-Moreno [b], Shayan Edalatmanesh[a,b], Koen Lauwaet[c], Michal Otyepka[a], Miroslav Medveď'[a], Álvaro Buendía[e], Rodolfo Miranda[c,e], Nazario Martín\*[c,d], Pavel Jelínek\*[a,b] and David Écija\*[c]**

\*Correspondence to: jelinekp@fzu.cz (P.J.), nazmar@ucm.es (N.M.), david.ecija@imdea.org (D.E.)

**This PDF file includes:**

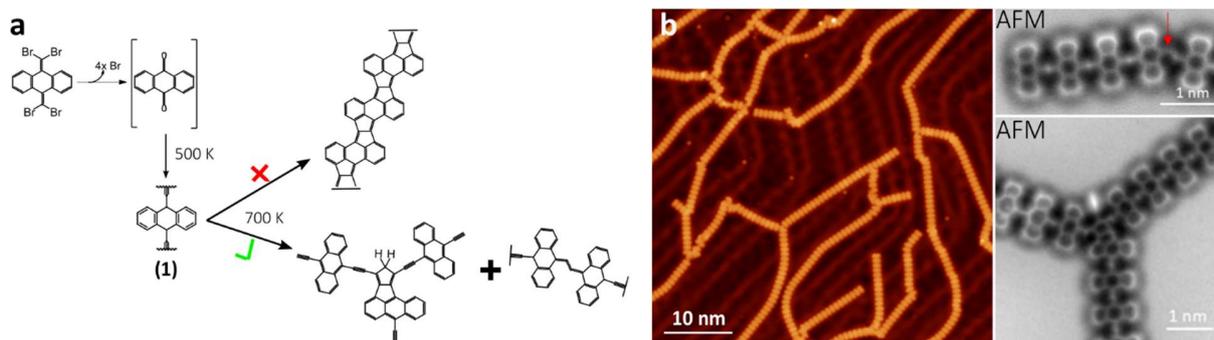

**Supplementary Fig. S1 | Polymerization evolution of 4BrAn precursor described in the main text. a,** Scheme of the reaction sequence of **4BrAn** precursor after being deposited on Au(111), annealed to 500 K to afford polymer **1**. Further annealing up to 700 K gives rise to inter-polymeric fusion, carbon atom addition and intra-polymeric warping instead of pentalene formation. **b,** STM overview (left panel) of **4BrAn** precursor after annealed to 700 K on Au(111), and AFM high-resolution detail of an intra-polymeric warping (top right panel) and inter-polymeric (right bottom panel) fused structures.

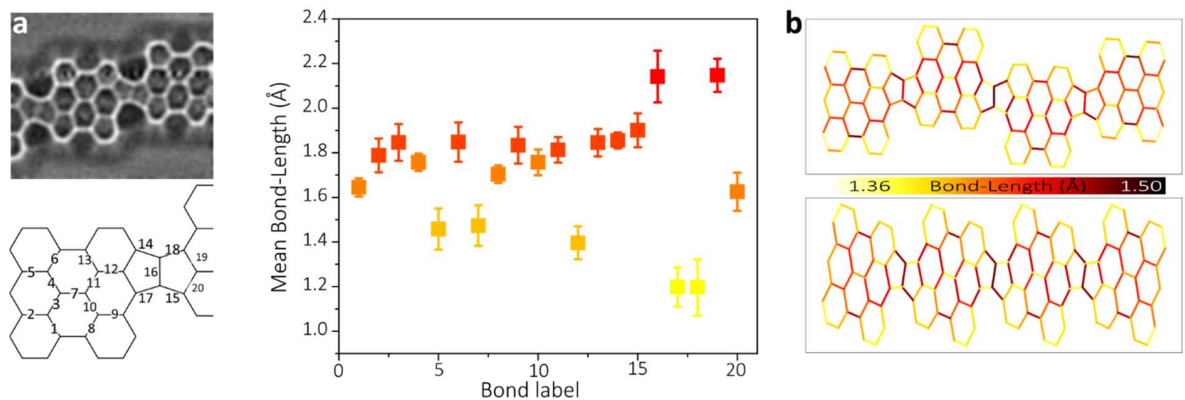

**Supplementary Fig. S2 | Experimental and theoretical bond-length analyses. a,** Typical nc-AFM high resolution image with applied Laplace-filtered to emphasise polymer structure used for obtaining experimentally C-C bond lengths (top-left panel) and C-C bond labelling (bottom left panel). Mean C-C bond-length plot obtained from a statistical analysis over several polymer structures. **b,** DFT calculated bond-length for *trans*- and *cis*- pentalene polymer displayed in a scale of colours.

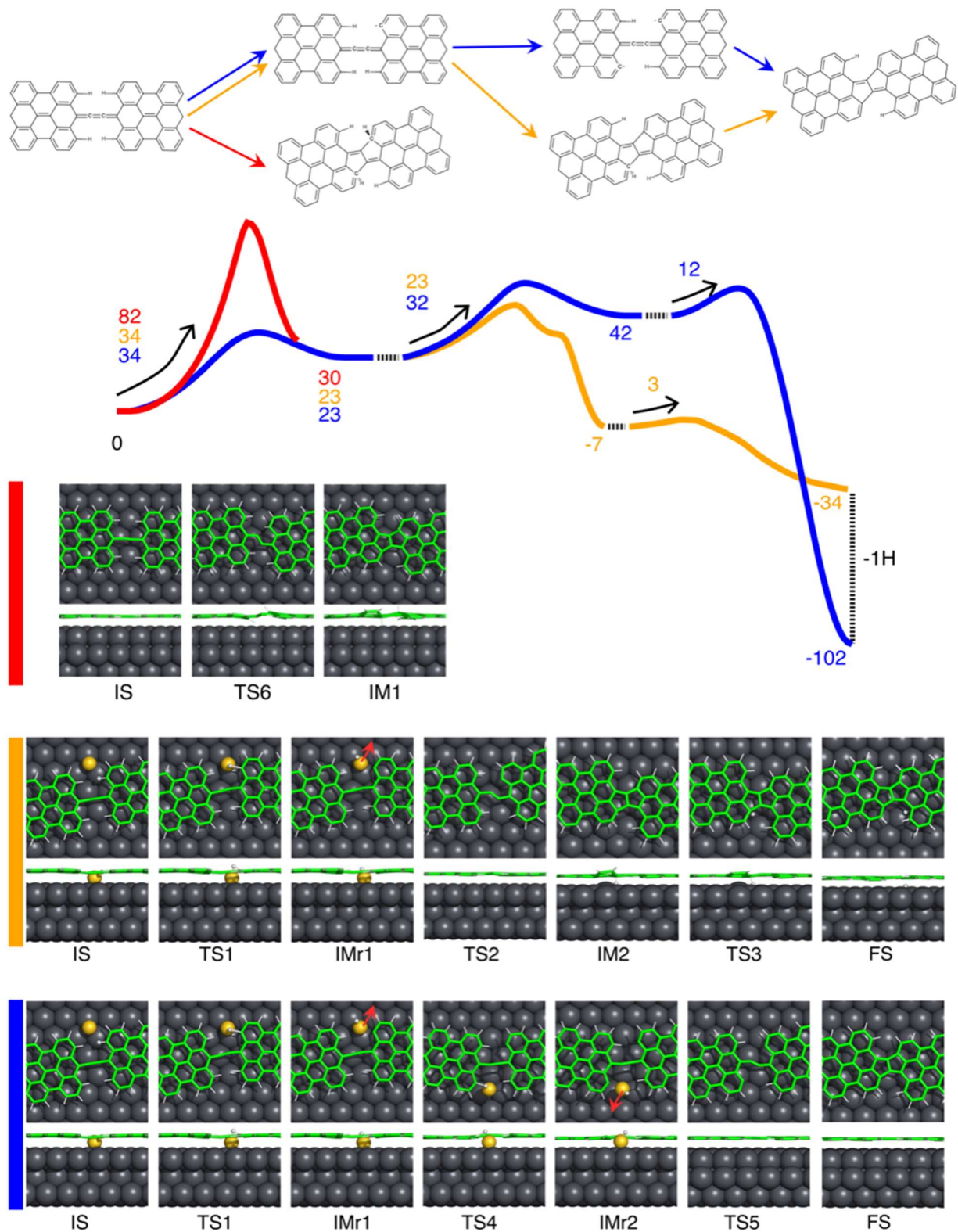

**Supplementary Fig. S3 | Scheme of calculated reaction pathways of bisanthene polymer (2).** Schematic view of distinct reaction pathways of transformation of cumulene-bridged bisanthene polymer to pentalene-bridged bisanthene

including energy values (kcal/mol) of initial (IS), transition (TS), intermediates (IM) and final (FS) states. Namely, we considered 3 possible reaction pathways: (i) direct cyclization (red); (ii) hydrogen dissociation, cyclization and hydrogen dissociation (yellow); and (iii) hydrogen dissociation, hydrogen dissociation and cyclization (blue). Corresponding atomic arrangements of different states along the reaction pathways are shown below.

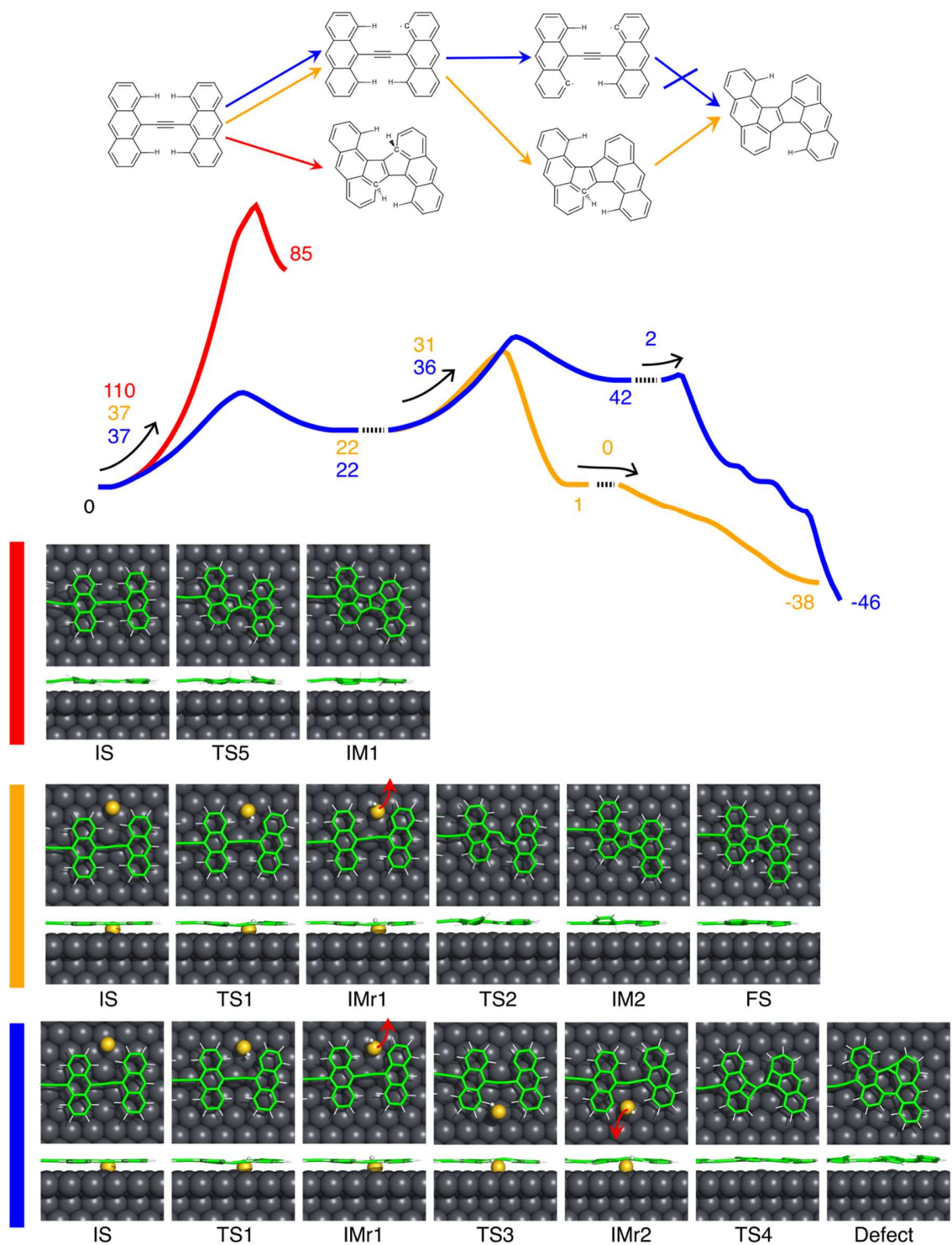

**Supplementary Fig. S4 | Scheme of calculated reaction pathways of anthracene polymer (1).** Schematic view of distinct reaction pathways of transformation of ethynylyne-bridged anthracene polymer to pentalene-bridged anthracene including energy values (kcal/mol) of initial (IS), transition (TS), intermediates (IM) and final (FS) states.

Namely, we considered 3 possible reaction pathways: (i) direct cyclization (red); (ii) hydrogen dissociation, cyclization and hydrogen dissociation (yellow); and (iii) hydrogen dissociation, hydrogen dissociation and cyclization (blue). Corresponding atomic arrangements of different states along the reaction pathways are shown below. Note that in the case of the dissociation-dissociation-cyclization reaction pathway, other structures instead of pentalene-bridged anthracene polymer are preferred.

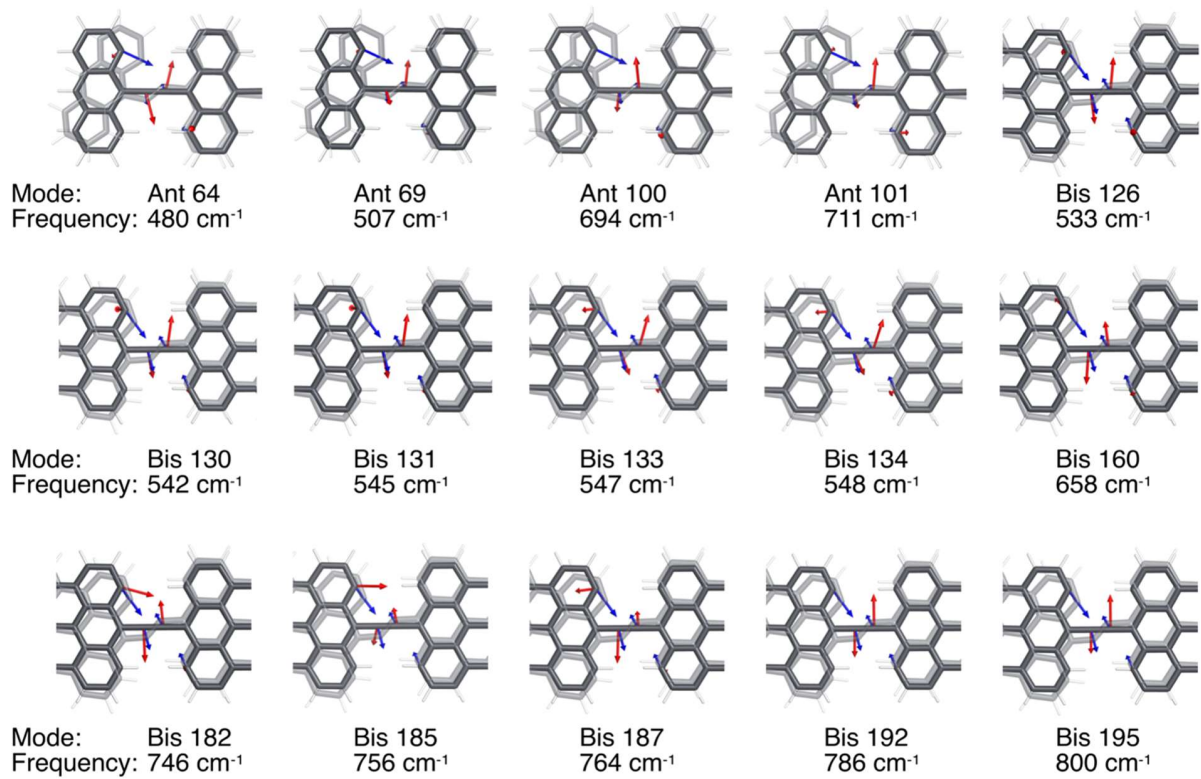

**Supplementary Fig. S5 | List of available bending vibrational modes in anthracene and bisanthene polymers and their comparison with the reaction coordinates.** We show all available vibrational modes with dominant contribution of the bending modes located on the bridging unit (corresponding vibrational eigenvectors are depicted by red arrows. Reaction coordinates (shown by blue arrows) are defined as the displacement from the initial (**IMr1**) to the transition state (**TS2**) for the cyclization reaction pathway from **IMr1** to **IM2**, see Supplementary Figs. 3 and 4. In the case of the anthracene polymer, none of the vibrational modes contains both the bending mode and characteristic in-phase motion of dehydrogenated carbon.